\documentclass[preprint,showpacs,showkeywords,preprintnumbers,amsmath,amssymb]{revtex4}

\usepackage{graphicx}
\usepackage{dcolumn}
\usepackage{bm}
\begin{document}

\title{Generalized Migdal-Kadanoff Bond-moving Renormalization Recursion Procedure II: Symmetrical Half-length Bond Operation on Fractals}

\author{Chun-Yang Wang\footnote{Corresponding author. Electronic mail:
wchy@mail.bnu.edu.cn}\footnote{Researcher in the Physical
Post-doctoral Circulation Station of Qufu Normal University}}
\author {Wen-Xian Yang}
\author {Zhi-Wei Yan}
\author {Hong Du}
\author {Xiang-Mu Kong}
\author {Yu-Qi Zhang}
\author {Ling-Yu Zhang}

\affiliation{Shandong Provincial Key Laboratory of Laser
Polarization and Information Technology, College of Physics and
Engineering, Qufu Normal University, Qufu 273165, China}


\begin{abstract}

In this second part of the series of two papers we report another
type of generalized Migdal-Kadanoff bond-moving renormalization
group transformation recursion procedures considering symmetrical
single bond operations on fractals. The critical behavior of the
spin-continuous Gaussian model constructed on the Sierpinski gaskets
is studied as an example to reveal its predominance in application.
Results obtained by this means are found to be in good conformity
with those obtained from other studies.

\end{abstract}



\maketitle

\section{INTRODUCTION}

As has been discussed in the first part of this series of two papers
(henceforth referred to as paper I), the Migdal-Kadanoff bond-moving
renormalization group transformation recursion procedure and its
extensions are very powerful for the study of near-critical
properties of various classical lattice systems with globally
symmetries \cite{Migdal1,Migdal2,Kadanoff}. However inconveniences
can still be found in treating some dual symmetrical systems such as
the fractals even if the generalized procedures presented in paper I
are used \cite{Wilson,Balkan,Kramers,Wegner}.

From another point of view, the fractal systems with local
symmetries have attracted much attention in the study of phase
transition and critical phenomena since the pioneering works by
Gefen and co-workers \cite{Genfen1,Genfen2,Genfen3}. In the past few
decades great effort has been devoted to the investigation of such
typical fractal systems as Koch-type curves \cite{Kongxm1},
diamond-type hierarchical lattices
\cite{Kongxm2,pinhost,linzqky,linzqea}, Bethe-type lattices
\cite{hanbchs} as well as Sierpinski carpets and gaskets
\cite{zqlin,sliyang}. The method of decimation
\cite{Kadanoff2,Nelson}, block transformation
\cite{Kadanoff3,Neimeijer,Nauenberg,Kadanoff4,Kadanoff5,Bell} and
cumulate expansion \cite{lijkongxm} are the usual means for
scientists to rely on. Yet the applying of bond-moving procedures on
these fractals systems is seldom reported.

We noticed the generalized bond-moving transformation recursion
procedures presented in paper I can be easily applied to the
Sierpinski gaskets under a simple alteration. Thus resulted in this
paper another type of generalization of the remarkable
Migdal-Kadanoff bond-moving renormalization group transformation. In
the following sections of this paper we will give in detail the
generalizations we have made on these procedures (Sec. \ref{sec2})
and their predominance in applications (Sec. \ref{sec3})
respectively by recurring them on the Sierpinski gaskets to
investigate the critical properties of the spin-continuous Gaussian
model. A summary of our conclusion and some further discussions are
presented in Sec. \ref{sec4}.

\section{generalization}\label{sec2}

\begin{figure}
\includegraphics[scale=0.9]{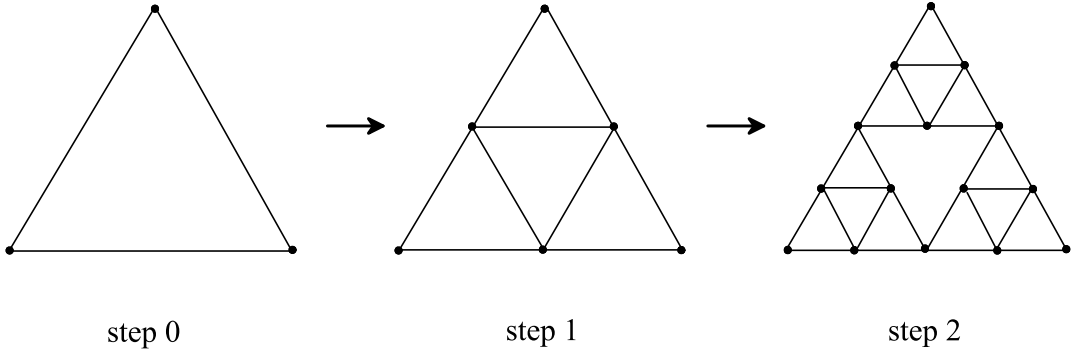}
\caption{Recursion procedures for the construction of Sierpinski
gaskets.\label{Sierlatt}}
\end{figure}

Sierpinski gaskets are a kind of typical infinitely ramified regular
fractal lattices that can be easily constructed by a repeated
process. The starting point is usually based on a regular triangular
as is illustrated in Fig. \ref{Sierlatt}a. Then connecting each
midpoint of the three sides to divide the initial triangular to
$B^{2}$ ($B=2$ here in this paper) smaller sub-triangulares, out of
which $l^{2}$ ($l=1$) sub-triangulares at the center of the initial
triangular is eliminated (Fig. \ref{Sierlatt}b). This procedure is
then infinitely repeated in the remained smaller sub-triangulares
(Fig. \ref{Sierlatt}c) iterating to the microscopic length scales.
The fractal dimension of this Sierpinski gaskets is then determined
by
\begin {equation}
D_{\textrm{f}}=\textrm{ln}(B^{2}-l^{2})/\textrm{ln}B=\textrm{ln}3/\textrm{ln}2.
\end {equation}

Here we can found that the partial structure of the Sierpinski
gaskets is actually still a triangular. Then we can deduce reversely
from this fact that it will be very easy to make the Sierpinski
gaskets coarse-grained if one proceeds following a very similar
renormalization procedure as those we have presented on the
triangular lattices in paper I.

\begin{figure}
\includegraphics[scale=0.8]{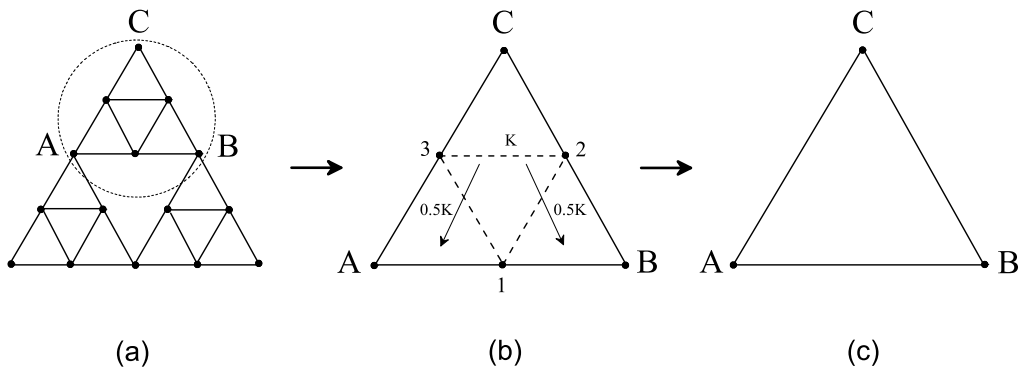}
\caption{Renormalization recursion procedures for the Sierpinski
gaskets to make the lattice coarse-grained in which symmetrical
half-length bond operations are considered. \label{tra-pro}}
\end{figure}

Basing on these considerations, we present here in the second part
of this series of two papers another type of generalized bond-moving
recursion procedures that can be used very conveniently on the
Sierpinski gaskets like fractal lattices. It proceeds in such a
little different way: (1) selecting a cluster of six lattice sites
in a small sub-triangular part of the Sierpinski gaskets as a basic
unit for recursion; (2) moving the to be eliminated bonds connecting
the three to be eliminated sites in the selected triplet with a
weight of half length bonds to the peripheral bonds. For an example
bond $2-3$ in $\triangle ABC$ connecting sites 2 and 3 is moved with
a half length weight to bonds $A-1$ and $1-B$ respectively as is
shown in Fig.\ref{tra-pro}b; (3) rescaling the system and decimating
the to be eliminated sites. Thus the lattice is made coarse-grained
under enough steps of renormalization. This procedure can also be
proved to be a powerful way bringing with great convenience in
particular in the study of spin-continuous systems constructed on
the fractal-like lattices.

\section{predominance illustration} \label{sec3}

For comparing and predominance illustrating of the above bond-moving
recursion procedures with other means, we study again the critical
behavior of the spin-continuous Gaussian model constructed on the
Sierpinski gaskets whose spins can take any real value between
$(-\infty,+\infty)$. The probability of finding a given spin between
$\sigma_{i}$ and $\sigma_{i} + d \sigma_{i}$ is assumed to be
$p(\sigma_{i})d\sigma_{i}\propto[\exp-(b/2)\sigma_{i}^{2}]
d\sigma_{i}$. This results in the classical Gaussian effective
Hamiltonian with two-body nearest-neighbor interactions
\begin {equation}
H_{\textrm{eff}}=\sum_{\langle
ij\rangle}K\sigma_{i}\sigma_{j}-\frac{b}{2}\sum_{i}\sigma_{i}^{2},
\end {equation}
where $K={J}/{k_{B}T}$ is the reduced nearest-neighbor interaction
with $K>0$ denotes the ferromagnetic systems; $b$ is the Gaussian
distribution constant; $k_{B}$ the Boltzmann constant and $T$ the
thermodynamic temperature. The summation of the spin is performed
between each nearest-neighbor pair $\langle ij\rangle$.

In order to successfully complete the bond-moving and decimation
processes and generalizing the Gaussian model on translational
invariant lattices to that on the Sierpinski gaskets, we assign two
types of interactions $K_{e}$ and $K$ for differentiation of spin
interaction at different cases as Gefen et al did in previous
studies \cite{Genfen3}. Where the bond noted as $K_{e}$ separates
between two to be eliminated sub-triangulares while $K$ borders a
non-eliminated one. For the particular case of Gaussian model two
types of self-energy $(-b_{e}s^{2}/2)$ and $(-bs^{2}/2)$ should also
be assigned correspondingly but the numerical value of $b_{e}$ and
$b$ is actually identical as well as that of $K_{e}$ and $K$. Thus
if a spin has $N$ bonds of $K$ and $N_{e}$ bonds of $K_{e}$ with its
nearest-neighbors, the self-energy of it is then given by
$[-(N_{e}b_{e}+Nb)s^{2}/2]$. In the renormalization procedures the
to be eliminated bonds are moved with a weight of half length to the
peripheral ones resulting a half weight addition to the
nearest-neighbor spin interaction but a wholly symmetrical
maintaining of the lattice.
\begin{figure}
\includegraphics[scale=1.1]{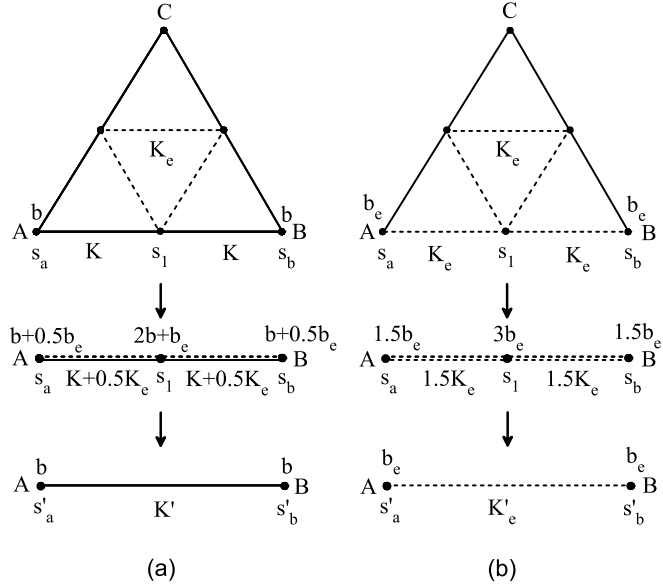}
\caption{Bond-moving and decimation processes to generalizing the
Gaussian model on translational invariant lattices to that on the
Sierpinski gaskets. \label{sl-pro}}
\end{figure}

As is illustrated in Fig.\ref{sl-pro}, along with the moving of the
nearest-neighbor interaction $Ks_{i}s_{j}$ (or $K_{e}s_{i}s_{j}$)
between the spins $s_{i}$ and $s_{j}$, the self-energy
$[-b(s_{i}^{2}+s_{j}^{2})/2]$ (or correspondingly
$[-b_{e}(s_{i}^{2}+s_{j}^{2})/2]$) move in the same direction. The
decimation procedure for the renormalized bond $K'$ is
\begin{eqnarray}
&&\int^{+\infty}_{-\infty}\textrm{exp}\left[\left(K+0.5K_{e}\right)\left(s_{a}s_{1}+s_{1}s_{b}\right)
-\frac{b+0.5b_{e}}{2}\left(s^{2}_{a}+s^{2}_{b}\right)-\frac{2b+b_{e}}{2}s^{2}_{1}
\right]ds_{1}
\nonumber\\&&\hspace{1.0cm}=C\textrm{exp}\left[K's'_{a}s'_{b}-\frac{b}{2}\left(s'^{2}_{a}+s'^{2}_{b}\right)\right]
\label{decimation}.
\end{eqnarray}
By directly integrating $s_{1}$ to decimate the intermediate spins
it becomes
\begin{eqnarray}
&&C\textrm{exp}\left[\left(\frac{(2K+K_{e})^{2}}{4(2b+b_{e})}\right)s_{a}s_{b}+\left(\frac{(2K+K_{e})^{2}}{16b+8b_{e}}
-\frac{2b+b_{e}}{4}\right)\left(s^{2}_{a}+s^{2}_{b}\right)\right]
\nonumber\\&&\hspace{1.0cm}=C\textrm{exp}\left[K's'_{a}s'_{b}-\frac{b}{2}\left(s'^{2}_{a}+s'^{2}_{b}\right)\right]
\label{decimation}.
\end{eqnarray}
For the continuity of spin sampling, the spins are rescaled by
\begin{eqnarray}
s'_{a}=\xi_{a}s_{a} \hspace{0.8cm} \textrm{and} \hspace{0.8cm}
s'_{b}=\xi_{b}s_{b} \label{rescale}
\end{eqnarray}
with
\begin{eqnarray}
\xi^{2}_{a}=\xi^{2}_{b}=1+\frac{b_{e}}{2b}-\frac{(2K+K_{e})^{2}}{4b(2b+b_{e})}\label{rescalefactor}.
\end{eqnarray}
Then the recursion relation for $K'$ is obtained to be
\begin{eqnarray}
K'=R(K,K_{e})=\frac{b(2K+K_{e})^{2}}{2(2b+b_{e})^{2}-(2K+K_{e})^{2}}\label{rescalefactor}.
\end{eqnarray}
Meanwhile, The renormalization group transformation of the cell
decimated to the renormalized bond $K_{e}'$ is
\begin{eqnarray}
&&\int^{+\infty}_{-\infty}\textrm{exp}\left[\frac{3}{2}K_{e}\left(s_{a}s_{1}+s_{1}s_{b}\right)
-\frac{3b_{e}}{4}\left(s^{2}_{a}+s^{2}_{b}\right)-\frac{3b_{e}}{2}s^{2}_{1}
\right]ds_{1}\nonumber\\&&\hspace{1.0cm}=C\textrm{exp}\left[\frac{3K_{e}^{2}}{4b_{e}}s_{a}s_{b}-\frac{3(2b_{e}^{2}-K_{e}^{2})}{8b_{e}}\left(s^{2}_{a}+s^{2}_{b}\right)\right]
\nonumber\\&&\hspace{1.0cm}=C\textrm{exp}\left[K'_{e}s'_{a}s'_{b}-\frac{b_{e}}{2}\left(s'^{2}_{a}+s'^{2}_{b}\right)\right]
\label{decimation},
\end{eqnarray}
from which the recursion relation for $K'_{e}$ is derived to be
\begin{eqnarray}
K'_{e}=R_{e}(K,K_{e})=\frac{b_{e}K_{e}^{2}}{2b_{e}^{2}-K_{e}^{2}}\label{rescalefactor}.
\end{eqnarray}

From these recursion relations we can found the critical behavior of
the Gaussian model on the Sierpinski gaskets is quite different from
that on the translational invariant lattices. Here we can obtain an
attractive fixed point $(K^{\ast}=0,K_{e}^{\ast}=0)$ and two
critical points $(K^{\ast}=b,K_{e}^{\ast}=0)$ and
$(K^{\ast}=0,K_{e}^{\ast}=b_{e})$. An fixed point
$(K^{\ast}=b,K_{e}^{\ast}=b_{e})$ was found to be repulsive
corresponding to the critical point of the Gaussian model on the
translational invariant lattices given $K_{e}=K$ and $b_{e}=b$.

Since the integration of the partition function are kept limited
because of the introduction of self-energies, we can obtain the
renormalization-group transformation matrix at
$(K^{\ast}=b,K_{e}^{\ast}=0)$ as
\begin{eqnarray}
R(K,K_{e})&=&\left(
\begin{array}{cc}
\frac{\partial K'}{\partial K} & \frac{\partial K'}{\partial
K_{e}} \\
\frac{\partial K'_{e}}{\partial K} & \frac{\partial K'_{e}}{\partial
K_{e}} \\
\end{array}
\right)_{(K^{*}=b,K_{e}^{*}=0)}\nonumber\\& =&\left(
\begin{array}{cc}
4 & 2 \\
0 & 0 \\
\end{array}
\right)
 \label{matrixa},
\end{eqnarray}
by setting $b_{e}=0$ at critical point
$(K^{\ast}=b,K_{e}^{\ast}=0)$. Clearly we can found that it has only
two eigenvalues $\lambda_{1}=4$ and $\lambda_{2}=0$. Thus we obtain
the critical exponent of correlation length as
\begin{eqnarray}
\nu=\frac{\textrm{ln}B}{\textrm{ln}\lambda_{1}}=\frac{\textrm{ln}2}{\textrm{ln}4}=0.5\label{criticalE},
\end{eqnarray}
in good conformity with the previous results
\cite{zqlin,chyw1,chyw2}.

In the meantime, the renormalization-group transformation matrix at
$(K^{\ast}=0,K_{e}^{\ast}=b_{e})$ is found to be
\begin{eqnarray}
R_{e}(K,K_{e})&=&\left(
             \begin{array}{cc}
               \frac{\partial K'}{\partial
K} & \frac{\partial K'}{\partial
K_{e}} \\
               \frac{\partial K'_{e}}{\partial
K} & \frac{\partial K'_{e}}{\partial
K_{e}} \\
             \end{array}
           \right)_{(K^{*}=0,K_{e}^{*}=b_{e})}\nonumber\\&
=&\left(
\begin{array}{cc}
0 & 0 \\
0 & 4 \\
\end{array}
\right)
 \label{matrixa},
\end{eqnarray}
by setting $b=0$. It has also two eigenvalues $\lambda_{1}=0$ and
$\lambda_{2}=4$ resulting in the correlation length critical
exponent
\begin{eqnarray}
\nu_{e}=\frac{\textrm{ln}B}{\textrm{ln}\lambda_{2}}=\frac{\textrm{ln}2}{\textrm{ln}4}=0.5\label{criticalE}
\end{eqnarray}
conforming to the previous results.

\section{summary and discussion} \label{sec4}

In summary, in this second part of the series of two papers we have
generalized the Migdal-Kadanoff bond-moving renormalization group
transformation recursion procedures to containing symmetrical single
bond operations that can be used conveniently in particular on the
fractal lattices. The critical behavior of the classical
spin-continuous Gaussian model constructed on the Sierpinski gaskets
was studied as an example of the application of these procedures.
Results obtained are in good conformity with previous studies
revealing the dependability of these means. The predominance of
these procedures revealed in this paper may encourage many future
applications of them on some more complicated spin systems such as
the $S^{4}$ models.

\section * {ACKNOWLEDGEMENTS}

This work was supported by the Shandong Province Science Foundation
for Youths (Grant No.ZR2011AQ016), the  Shandong Province
Postdoctoral Innovation Program Foundation (Grant No.201002015), the
Scientific Research Starting Foundation, Youth Foundation (Grant
No.XJ201009) and the Foundation of Scientific Research Training Plan
for Undergraduate Students (Grant No.2010A023) of Qufu Normal
University.

\end{document}